\documentstyle[aps,prl,epsfig]{revtex}

\begin{document}
\draft	

\twocolumn[\hsize\textwidth\columnwidth\hsize\csname@twocolumnfalse\endcsname

\title{Fluctuation-dissipation ratio in lattice-gas models with kinetic 
constraints}

\author{Mauro Sellitto}

\address{Dipartimento di Scienze Fisiche and Unit\`a INFM, \\
Universit\`a ``Federico II'', Mostra d'Oltremare, Pad.~19,
I-80125 Napoli, Italy \\
{\tt Mauro.Sellitto@na.infn.it} \\
{\rm and} \\
Instituto de F\'\i sica, UFRGS, CP 15051  \\
91501-970 Porto Alegre RS Brazil      
}

\date\today

\maketitle

\begin{abstract}
We investigate by Montecarlo simulation the linear response function 
of three dimensional structural glass models defined by short-range 
kinetic constraints and a trivial equilibrium Boltzmann-Gibbs measure.
The breakdown of the fluctuation-dissipation theorem
in the glassy phase follows the prediction of mean field low temperature 
mode-coupling theory.

\end{abstract}


\twocolumn\vskip.5pc]\narrowtext

\paragraph*{Introduction.}
Glassy materials are a paradigmatic example of systems characterized by slow 
relaxation  and physical aging phenomena~\cite{Struik,Zar}.
While thermodynamic observables are rather regular along the glass 
transformation, dynamical properties undergo drastic changes, 
and in particular relaxation times increases up to 15 orders of magnitude.
The microscopic origin of such a dramatic slowing down
is the so called {\it cage effect\/}: a tagged 
particle in a liquid does not perform a simple random walk but 
rattles in a cage formed by the surrounding particles, and moves
away from it only if the particles building up the cage walls 
are themselves able to move.
At low temperature or high density, when the degrees of freedom of 
the particles become strongly coupled, this cooperative process
exceeds observation time and the glass appears like a ``solid''~\cite{Gotze}. 

However, structural relaxation is not completely absent in glasses,
and molecular rearrangements  still occur on longer time scales.
Indeed, after a quench in the low temperature or high density phase,
particle diffusion does not stop altogether, but 
rather becomes slower and slower as time goes on.
This is the origin of {\it aging effect\/}:
the response to an external perturbation depends on the ``age'' of the
sample, i.e., on the time elapsed since its preparation, 
even at very long times. 
Therefore in this regime glassy systems never reach thermodynamic
equilibrium and, consequently, time translational invariance (TTI) 
and the fluctuation-dissipation theorem (FDT) are violated.

\paragraph*{The problem.}
Despite the remarkable progress realized in the study of out-of-equilibrium 
dynamics of aging systems~\cite{review},
the detailed nature of ergodicity breaking in finite-dimensional glasses
is not yet fully understood.
In mean-field mode-coupling theory the glass transition appears as a purely 
dynamic effect due to an instability of the equation governing the correlation 
of density fluctuations~\cite{Gotze}. 
In particular, mean-field disordered models of structural glasses 
show that the origin of this dynamical transition is the existence of a 
large number of metastable states which trap the system for an 
infinite time.
On the other hand, the life-time of metastable states in finite-dimensional 
short range models is finite, since it is always possible to nucleate, by a 
thermally activated process, a droplet of the stable phase.
Therefore the dynamical transition appear as an artifact 
of the mean-field approximation, and in real glasses this transition 
would be a {\it finite-time\/} effect, at least on time scales much 
smaller than the life-time of metastable states~\cite{Das,Kirk,Remi,FrPa}.

However, since 
salient features of glassy behavior are essentially of dynamical nature, 
and since dynamical universality classes are smaller 
than the static ones, it is important to establish how glassy effects 
depend on the details of dynamics.
In this paper we explore the limit case of three dimensional 
lattice gas models defined only by short-range kinetic constraints and 
by a trivial equilibrium measure, in which the glass transition appears to 
have a purely dynamical origin unrelated to the existence of metastable
states~\cite{KA,KuPeSe}.
Previous Montecarlo simulations have shown that this model reproduces
qualitatively the ``glassy" phenomenology, 
including annealing-rate dependence, irreversibility effects,  
and waiting-time dependent logarithmic diffusion~\cite{KuPeSe,PeSe}.
Here we investigate aging effects in the linear response function in order 
to characterize the violation of FDT~\cite{CuKu1,CuKuPe}.

\paragraph*{The model.}
Our starting point is the kinetic lattice-gas model introduced 
by Kob and Andersen in order to mimic the cage effect in 
supercooled liquids~\cite{KA}.
The system consists of $N$ particles in a cubic lattice of size $L^{3}$, 
with periodic boundary conditions.
There can be at most one particle per site. 
Apart from this hard-core constraint there are no other static 
interactions among the particles.
At each time step a particle and one of its neighbouring sites are 
chosen at random. 
The particle moves if the three following conditions are all met:
\begin{enumerate}
\item the neighbouring site is empty;
\item the particle has less than $4$ nearest neighbours;
\item the particle will have less than $4$ nearest neighbours after 
it has moved.
\end{enumerate}
The rule is symmetric in time, detailed balance is satisfied and the 
allowed configurations have the same weight in equilibrium. 
With this simple definition one can proceed to study the dynamical 
behavior of the model at equilibrium. 
One observes that the dynamics becomes slower and slower as the 
particle density $\rho$ increases; in particular, the diffusion 
coefficient of the particles, $D$, vanishes as
the density $\rho$ approaches the
critical value $\rho_{\rm c}\simeq 0.88$, with a power law 
\begin{eqnarray}
D(\rho) & \sim & (\rho_c-\rho)^{\phi}, 
\end{eqnarray}
with an exponent $\phi \simeq 3.1$~\cite{KA}.

Since we are interested in the dynamical approach to the putative 
equilibrium state we allow the system to exchange particles with a
reservoir characterized by a chemical potential $\mu$.
Therefore, we alternate the ordinary diffusion sweeps with
sweeps of creation/destruction of particles on a single layer 
with the following Montecarlo rule: we randomly choose a site on 
the layer; if it is empty, we add a new particle; otherwise we 
remove the old particle with probability $\mbox{e}^{-\beta \mu}$.
The number of particles is no longer fixed and the external control 
parameter is $\mu$, which plays the role of the inverse temperature.
In this way we can prepare the system in a non equilibrium state
by a process analogous to a quench,
which is represented by a jump in  $1/\mu$ from above to below
$1/\mu_c$ (where $\mu_c$ is defined through the state equation
$\rho(\mu_c)=\rho_c$). 
The situation becomes analogous to the {\it canonic\/} case in 
which one controls the temperature, and the energy endeavors to reach  
its equilibrium value.

\paragraph*{Linear response.} 
Following a suggestion by J. Kurchan,  we compute the linear response function
by applying to our system a small random perturbation at time $t_w$:
\begin{eqnarray}
{\cal H}_{\epsilon} &=&
\epsilon \sum_{a=1}^3 \sum_{k=1}^N f^a_k \cdot r^a_k \,\,,
\end{eqnarray}
where $ f^a_k=\pm1$ independently for each coordinate $a$ and
particle $k$ at position $r^a_k$.
The response function of the system at time $t+t_w$ is defined by
\begin{eqnarray}
	R(t+t_w,t_w) 
&=& 	
	\frac{1}{3N} \sum_{a=1}^3 \sum_{k=1}^N \left. \left\langle
	\frac{\partial f_k^a \cdot r_k^a(t+t_w)}{\partial \epsilon(t_w)} 
	\right\rangle \right|_{\epsilon=0} .
\end{eqnarray}					
The linear regime is probed for small enough  values of the perturbation
strength $\epsilon$. 
It is actually more convenient to look at the integrated response function:
\begin{eqnarray}
\chi(t+t_w,t_w) &=&
\int_{t_w}^{t+t_w} R(t+t_w,\tau) \epsilon (\tau) d\tau \,\,.
\end{eqnarray}
In order to measure this quantity we prepare the system in a non-equilibrium 
state by quenching to the subcritical value of $1/\mu$ and letting it 
relax up to time $t_w$; 
at $t_w$, we make a copy of the system and apply
the perturbation; we then let evolve the two systems with the 
{\em same\/} succession of random numbers and  measure the 
difference $\Delta r_k^a(t+t_w)$ between the displacements that take place 
in the two systems at time $t+t_w$.
At constant field we obtain
\begin{eqnarray}
\chi(t+t_w,t_w) &=&
\frac{1}{3N} \sum_{a=1}^3 \sum_{k=1}^N 
\left\langle f_k^a \cdot \Delta r_k^a(t+t_w) \right\rangle.
\end{eqnarray}
 
We have performed a sudden quench to the subcritical value 
$1/\mu=1/2.2$ starting from a configuration with density $0.75$ 
and measured the integrated response function $\chi(t+t_w,t_w)$ 
vs.\ time $t$ at different waiting times $t_w$. 
We have checked that for  $0.05 < \epsilon< 0.15$  non-linear effects 
are absent.
In the following we present the results for the case $\epsilon=0.1$
and for a cubic lattice of size $20^3$.
Figure \ref{chi} shows that for small values of $t_w$ TTI holds, and
aging effects set in only for larger value of $t_w$.
To obtain the scaling behavior one has to consider an effective waiting
time, $\tau_w=t_w+\tau_0$, that takes into account the relaxation time, 
$\tau_0$, of the system before the quench.
Indeed, figure \ref{chi_scaled} shows that, if one plots 
$\chi(t+t_w,t_w)$ vs. $t/\tau_w$, the curves lie roughly
on top of each other and follow a {\it simple\/} aging logarithmic 
behavior, $\chi(t+t_w,t_w) \sim \log(1+t/\tau_w)$.
Similar results were previously found also for the mean square 
displacement~\cite{KuPeSe,PeSe}.

\paragraph*{Fluctuation-dissipation ratio.}
We are now in a position to  investigate the nature of breakdown
of thermodynamical equilibrium.
We consider the generalized FDT proposed by Cugliandolo and Kurchan in 
the framework of dynamical mean-field theory of spherical $p$-spin glass 
model~\cite{CuKu1}.
In our case the relevant property is the off-equilibrium
generalization of Einstein-Stokes relation among diffusion coefficient 
and viscosity, given by~\cite{CuLe}:
\begin{eqnarray}
 X(B) \frac{\partial B(t+t_w,t_w)}{\partial t_w}=-2 T R(t+t_w,t_w),
\end{eqnarray}								
where 
\begin{eqnarray}
 B(t+t_w,t_w)= \frac{1}{3N} \sum_{a=1}^3 \sum_{k=1}^N 
\left\langle \left[r_k^a(t+t_w)-r_k^a(t_w) \right]^2 \right\rangle,
\end{eqnarray}
is the mean square displacement.
The fluctuation-dissipation ratio (FDR), $X$, depends on both times only 
through $B$ and its departure from 1 is a measure of violation of FDT.
The integrated version of the generalized FDT allows to write 
(considering a constant field):
\begin{eqnarray}
\chi(t+t_w,t_w) &=&
\frac{\epsilon}{2T}
\int^{B(t+t_w,t_w)}_0 X(B) \, dB \,\,.
\end{eqnarray}
(We always write explicitly the temperature $T$, however it does not
play any role.)
Therefore, if FDT holds, we would have a straight line with slope $\epsilon/2T$
in the parametric plot of $B(t+t_w,t_w)$ versus $\chi(t+t_w,t_w)$;
while a deviation from this straight line indicates a failure of FDT.
The way in which FDT breaks down plays a key role in aging systems 
since it signals the presence of a time-scale dependent effective 
temperature~\cite{CuKuPe}, and this violation  can be bounded by 
a quantity related to the entropy production~\cite{CuDeKu}. 
The FDR has been measured by 
Montecarlo simulations in several models such as  
spin-glasses~\cite{SilvioHeiko,Marinari}, 
Lennard-Jones binary mixtures~\cite{Giorgio},
domain growth models~\cite{Alain}, 
and directed polymers~\cite{Yoshino},
and so far all the findings are consistent with the prediction
of generalized FDT.  
We have carried out extensive numerical simulation to measure
the FDR, $X(B)$, in three dimensional lattice-gas models with 
kinetic constraints.
We observe the two asymptotic regimes also found in the out of 
equilibrium dynamics of mean field models of structural 
glasses~\cite{CuKu1}: 
\begin{itemize}
\item
at times $t$ smaller than $t_w$ we observe
a short-time quasi equilibrium regime where FDT holds;
\item
at larger separation times FDT breaks down
with a constant violation factor.
\end{itemize}
These results appear clearly in the parametric plot~\cite{CuKu2} of  
$\chi(t+t_w,t_w)$ vs.\  $B(t+t_w,t_w)$ where, as can be seen in  
fig.~\ref{x22}, the curve approaches the characteristic broken line.
(In order to probe the asymptotic time regime we have chosen $t_w=10^5$.) 
It is natural in this context to interpret the quasi-equilibrium regime
as corresponding to the motion of particle inside a cage, although the size 
of cage in this model is quite small and of the order of a fraction of 
lattice spacing. 
After that the system crosses over a diffusion-cage regime 
where a departure from equilibrium is observed with a constant
FDR.

In order to see if the inverse chemical potential plays the role
of a ``true'' temperature, we have measured the function $X(B)$  
by quenching the system at different subcritical values.
Figure~\ref{xmu} shows that the slope of the FDT violating line, 
as well as the location of the breaking point, are independent 
of the quench value of $1/\mu$, at least in the time window explored 
in our numerical experiments and in the range of values of $1/\mu$ used.
Therefore as suggested in ref.~\cite{PeSe}, the FDR appears locked in 
at the value the system would exhibit at the critical density, 
and should be only related with ``universal'' properties 
of the model, like for instance the exponent $\phi$.

\paragraph*{Conclusions.}
To summarize we have shown that three dimensional lattice-gas 
models defined by short range kinetic constraint and trivial 
equilibrium Boltzmann-Gibbs measure exhibit a {\it simple\/} aging 
behavior in the integrated response function and violate 
fluctuation-dissipation theorem in a manner similar to mean 
field models of structural glasses. 
The fluctuation-dissipation ratio in the glassy phase appears to be 
a constant independent on the chemical potential of the reservoir; 
however, different results are expected if the boundary condition is 
changed or if creation and destruction of particles within the bulk 
is allowed.

Given the non-holonomic nature of kinetic constraints and the trivial 
hamiltonian of the model, any statistical mechanics approach based on 
the calculation of some partition function should be unsuitable to grasp 
the features of the glassy phase.
In order to understand if aging is interrupted after a certain characteristic 
time~\cite{FrMePaPe}, it would be interesting to analytically investigate 
the long-time dynamical behavior of the model.
A persistent aging scenario would prove that also in finite-dimensional 
short-range models the glass transition may have a purely dynamic origin 
unrelated to any underlying thermodynamic singularity. 

\acknowledgements
This work was originated by inspiring collaboration with 
Jorge Kurchan and Luca Peliti. 
I wish to thank them for several key suggestions and 
discussions.
I am grateful to Jeferson J. Arenzon and Cl\'audia P. Nunes 
for the kind hospitality in Porto Alegre where this work was 
completed.

\begin{figure}[f]
\begin{center}
\epsfig{file=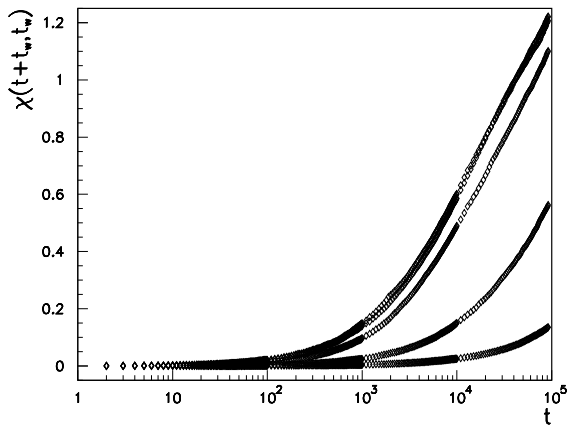,width=9cm} 
\caption{Integrated response function $\chi(t+t_w,t_w)$ vs. time $t$
after a quench to the subcritical values 
$1/\mu=1/2.2$. 
The measurements were carried out after the perturbation 
was applied at waiting times $t_w=10,10^2,10^3,10^4,10^5$.
The initial density is $0.75$.
Average over $20$ samples.}
\label{chi}
\end{center}
\end{figure}

\begin{figure}[f]
\begin{center}
\epsfig{file=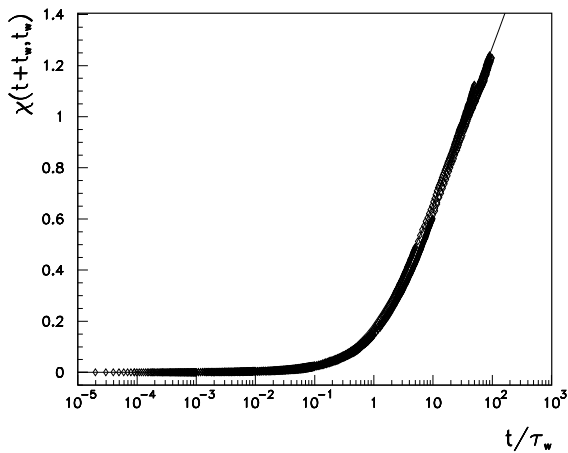,width=9cm} 
\caption{Scaling behavior of integrated response function 
$\chi(t+t_w,t_w)$ vs.\  $t/\tau_w$, where $\tau_w=t_w+\tau_0$
with $\tau_0=10^3$ and $t_w=10,10^2,10^3,10^4,10^5$.
The full line corresponds to $\chi(t+t_w,t_w) \sim \log(1+t/\tau_w)$.}
\label{chi_scaled}
\end{center}
\end{figure}

\begin{figure}[f]
\begin{center}
\epsfig{file=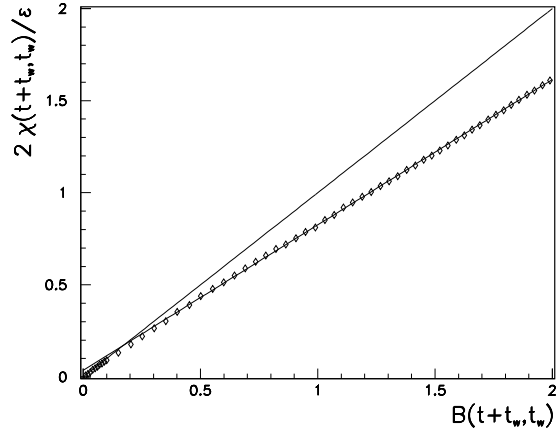,width=9cm} 
\caption{Parametric plot of  $2 \chi(t+t_w,t_w)/\epsilon$ 
vs.\ mean square displacement $B(t+t_w,t_w)$
measured after a waiting time $t_w=10^5$ from a quench to the 
subcritical value $1/\mu= 1/2.2$. 
The density of initial configuration is $0.75$.
The two straight line have slope $1$ (FDT line) and $0.79$.
Average over 200 samples; error bars
comparable to symbol size.}
\label{x22}
\end{center}
\end{figure}

\begin{figure}[f]
\begin{center}
\epsfig{file=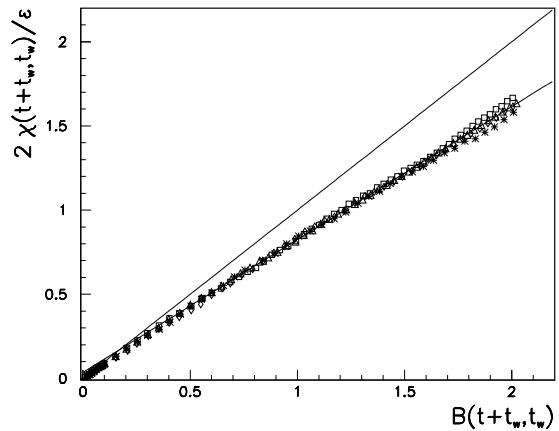,width=9cm} 
\caption{Parametric plot of  
$2 \chi(t+t_w,t_w)/\epsilon$ vs.\ mean square displacement $B(t+t_w,t_w)$
measured after a waiting time $t_w=10^5$ from a quench to the 
subcritical value $1/\mu= 
1/2.05$ (stars), $1/2.1$ (triangles), $1/2.2$ (diamonds), $1/2.3$ (squares). 
The density of initial configuration is $0.75$.
The two straight line have slope $1$ and $0.79$.
Average over 200 samples; error bars comparable to
symbols size.}
\label{xmu}
\end{center}
\end{figure}

\end{document}